\documentclass[review]{elsarticle}

\usepackage{lineno,hyperref,amsmath,amssymb}
\modulolinenumbers[5]

\journal{Journal of \LaTeX\ Templates}

\begin{document}

\begin{frontmatter}

\title{Delay reductions of the two-dimensional Toda lattice equation}

\author[1]{Aika Tsunematsu\corref{cor1}}

\author[1]{Kenta Nakata}

\author[1]{Yuta Tanaka}

\author[2]{Ken-ichi Maruno\corref{mycorrespondingauthor}}

\address[1]{Department of Pure and Applied Mathematics, School of Fundamental Science and Engineering, Waseda University, 3-4-1 Okubo, Shinjuku-ku, Tokyo 169-8555, Japan }
\address[2]{Department of Applied Mathematics, Faculty of Science and Engineering, Waseda University, 3-4-1 Okubo, Shinjuku-ku, Tokyo 169-8555, Japan}

\cortext[mycorrespondingauthor]{Corresponding author}
\ead{kmaruno@waseda.jp}

\begin{abstract}
Integrable delay analogues of the two-dimensional Toda lattice equation are presented 
and their muti-soliton solutions are constructed by applying the delay reduction to 
the Gram determinant solution. 
\end{abstract}


\end{frontmatter}

\linenumbers

\section{Introduction}

Delay-differential equations have been used as mathematical models 
in various fields of science and engineering, such as traffic flow and infectious diseases. 
Studies on integrability and exact solutions of delay-differential equations 
have been carried out from mathematical and applied point of view
\cite{hasebe,kanai,Quispel,Levi,grammaticos,
ramani,ablowitz,joshi1,joshi2,carstea,viallet,halburd,BK,Stokes}. 
Integrable delay-differential equations have been obtained by 
similarity reductions or symmetry reductions in previous studies, 
and it was confirmed that these integrable delay-differential equations exhibit 
singularity confinement type behavior which is one of important properties
of discrete integrable systems\cite{SC}. 
Villarroel and Ablowitz considered a delay analogue of the 
two-dimensional Toda lattice (2DTL) equation and found 
the Lax pair and established the inverse scattering transform\cite{ablowitz}. 
However multi-soliton solutions have not been known until now.  

In this article, we present integrable delay analogues of 
the 2DTL equation and construct their multi-soliton 
solutions by a delay reduction. 
As far as we know, 
this is the first time that multi-soliton solutions of 
delay-differential equations has been obtained. 

\section{Delay analogues of the 2DTL equation and their multi-soliton solutions}

We consider the 2DTL equation
\begin{equation}
\frac{\partial^2}{\partial x\partial y} \log(1+V_n)=V_{n+1}-2V_n+V_{n-1}\label{2DTL1}
\end{equation}
which can be written as
\begin{equation}
 \frac{\partial^2 r_n}{\partial x\partial y}=e^{r_{n+1}}-2e^{r_n}+e^{r_{n-1}}
\end{equation} 
or
\begin{equation}
 \frac{\partial^2 u_n}{\partial x\partial y}
=e^{u_{n-1}-u_n}-e^{u_{n}-u_{n+1}}
\end{equation} 
via the dependent variable transformation $r_n=u_n-u_{n+1}=\log(1+V_n)$. 
The 2DTL equation \eqref{2DTL1} is transformed into the bilinear equation 
\begin{equation}
D_xD_y\tau_n\cdot \tau_n=2(\tau_{n+1}\tau_{n-1}-\tau_n^2)
\end{equation}
via the dependent variable transformation
\begin{equation}
V_n(x,y)=\frac{\partial^2}{\partial x\partial y} \log \tau_n
=\frac{\tau_{n+1}\tau_{n-1}}{\tau_n^2}-1\,.
\end{equation}
Here $D_x$ and $D_y$ are Hirota's $D$-operator which is defined as 
\begin{equation}
D_x^mD_y^nf\cdot g=(\partial_x-\partial_{x'})^m(\partial_y-\partial_{y'})^nf(x,y)g(x',y')\vert_{x=x',y=y'}\,.
\end{equation}
The Gram determinant form of the $N$-soliton solution to the 2DTL equation \eqref{2DTL1} 
is given as follows\cite{Hirota}:
\begin{eqnarray}
&&V_n(x,y)=\frac{\partial^2}{\partial x\partial y} \log \tau_n
=\frac{\tau_{n+1}\tau_{n-1}}{\tau_n^2}-1\,,\label{2DTL1-sol}\\
&&\tau_n=\det \left(\delta_{ij}+\frac{\phi_i \psi_j}{p_i-q_j}\right)_{_{1\leq i,j\leq N}}\nonumber\\
&&\quad =\frac{\prod_{i=1}^N \phi_i}{\prod_{j=1}^{N} \phi_j}
\det \left(\delta_{ij}\frac{\phi_j}{\phi_i}+\frac{\phi_j \psi_j}
{p_i-q_j}\right)_{_{1\leq i,j\leq N}}\nonumber\\
&&\quad     
=
\det \left(\delta_{ij}+\frac{\phi_j \psi_j}{p_i-q_j}\right)_{_{1\leq i,j\leq N}},\nonumber\\
     &&\phi_i=
      p_i^ne^{p_ix-p_i^{-1}y+\phi_i^{(0)}}\,, \quad
       \psi_i =q_i^{-n}
      e^{-q_ix+q_i^{-1}y+\psi_i^{(0)}}\,, \nonumber \\
&& \quad (i=1,2,\cdots,N)\,,\nonumber
\end{eqnarray}
where $p_i, q_i$ are positive constants and 
$\phi_i^{(0)}, \psi_i^{(0)}$ are real constants. 

Applying the delay reduction 
\begin{equation}
V_n(x,y)=w(z,y)\,, \qquad z=x+hn\,,
\end{equation}
to the 2DTL equation \eqref{2DTL1}, 
where $h$ is a nonzero real constant, 
we obtain the delay 2DTL equation
\begin{equation}
\frac{\partial^2}{\partial z\partial y}\log(1+w(z,y))
=w(z+h,y)-2w(z,y)+w(z-h,y)\,,\label{delay2dtoda1}
\end{equation}
which can be written as
\begin{equation}
 \frac{\partial^2 r(z,y)}{\partial z\partial y}=e^{r(z+h,y)}-2e^{r(z,y)}+e^{r(z-h,y)}
\end{equation} 
or
\begin{equation}
 \frac{\partial^2 u(z,y)}{\partial z\partial y}
=e^{u(z-h,y)-u(z,y)}-e^{u(z,y)-u(z+h,y)}
\end{equation} 
via the dependent variable transformation $r(z,y)=u(z,y)-u(z+h,y)=\log(1+w(z,y))$. 
Note that this delay 2DTL equation is a delay partial differential equation.  
The delay 2DTL equation \eqref{delay2dtoda1} is transformed into the delay bilinear equation
\begin{equation}
D_zD_yf(z,y)\cdot f(z,y)=2(f(z+h,y)f(z-h,y)-f(z,y)^2)
\end{equation}
via the dependent variable transformation
\begin{equation}
w(z,y)=\frac{\partial^2}{\partial z\partial y} \log f(z,y)
=\frac{f(z+h,y)f(z-h,y)}{f(z,y)^2}-1\,.
\end{equation}

We impose the reduction condition
\begin{equation}
 \log p_i- \log q_i=h(p_i-q_i)\label{delay2dtodacond1}
\end{equation}
to the parameters $p_i, q_i$ $(i=1,2, \cdots , N)$ in the $N$-soliton solution \eqref{2DTL1-sol}.
Note that the reduction condition \eqref{delay2dtodacond1} 
includes the delay parameter $h$. 
Then, by setting $z=x+hn$, we have
\begin{eqnarray}
\phi_j\psi_j&=&p_j^ne^{p_jx-p_j^{-1}y+\phi_j^{(0)}}q_j^{-n}
      e^{-q_jx+q_j^{-1}y+\psi_j^{(0)}}\nonumber\\
&=&e^{p_jx-p_j^{-1}y+n\log p_j+\phi_j^{(0)}}e^{-q_jx+q_j^{-1}y-n\log q_j+\psi_j^{(0)}}\nonumber\\
&=&e^{(p_j-q_j)x+n(\log p_j-\log q_j)-(p_j^{-1}-q_j^{-1})y+\phi_j^{(0)}+\psi_j^{(0)}}\nonumber\\
&=&e^{(p_j-q_j)x+(p_j-q_j)hn-(p_j^{-1}-q_j^{-1})y+\phi_j^{(0)}+\psi_j^{(0)}}\nonumber\\
&=&e^{(p_j-q_j)(x+hn)-(p_j^{-1}-q_j^{-1})y+\phi_j^{(0)}+\psi_j^{(0)}}\nonumber\\
&=&e^{(p_j-q_j)z-(p_j^{-1}-q_j^{-1})y+\phi_j^{(0)}+\psi_j^{(0)}}\,.
\end{eqnarray}
Thus, by imposing the reduction condition \eqref{delay2dtodacond1} and setting $z=x+hn$, 
$f(z,y)=\tau_n(x,y)$, 
we obtain the following $N$-soliton solution of the delay 2DTL equation \eqref{delay2dtoda1}:
\begin{eqnarray}
&&w(z,y)=\frac{\partial^2}{\partial z\partial y} \log f(z,y)\nonumber\\
&&\hspace{1.0cm} =\frac{f(z+h,y)f(z-h,y)}{f(z,y)^2}-1
\,,\\
 && f(z,y)
=\det \left(\delta_{ij}+\frac{\Phi_j(z,y)}{p_i-q_j}\right)_{_{1\leq i,j\leq N}}\,,
\nonumber\\
     &&\Phi_j(z,y)=e^{(p_j-q_j)z-(p_j^{-1}-q_j^{-1})y+\Phi_j^{(0)}}\,,\nonumber
\end{eqnarray}
where $p_i, q_i$ must satisfy 
\begin{equation}
\log p_i- \log q_i=h(p_i-q_i)\,, \quad
       (i=1,2,\cdots,N)
\end{equation}
and $\Phi_i^{(0)}$ are real constants.

Next, we consider the delay reduction of the 2DTL equation in a different coordinate and 
its one-dimensional reduction.  
By the independent variable transformation
\begin{equation}
x=\frac{t+s}{2}\,,\quad y=\frac{t-s}{2}\,,
\end{equation}
the 2DTL equation \eqref{2DTL1} leads to 
\begin{equation}
\left(\frac{\partial^2}{\partial t^2}-\frac{\partial^2}{\partial s^2}\right) 
\log(1+V_n(t,s))=V_{n+1}(t,s)-2V_n(t,s)+V_{n-1}(t,s)\,,\label{2DTL2}
\end{equation}
which can be written as
\begin{equation}
 \frac{\partial^2 r_n}{\partial t^2}-\frac{\partial^2 r_n}{\partial y^2}
=e^{r_{n+1}}-2e^{r_n}+e^{r_{n-1}}\,,
\end{equation} 
or
\begin{equation}
\frac{\partial^2 u_n}{\partial t^2}-\frac{\partial^2 u_n}{\partial y^2}
=e^{u_{n-1}-u_n}-e^{u_{n}-u_{n+1}}\,.
\end{equation} 
via the dependent variable transformation $r_n=u_n-u_{n+1}=\log(1+V_n)$. 
The $N$-soliton solution of the 2DTL equation \eqref{2DTL2}
is given as follows:
\begin{eqnarray}
&&V_n(t,s)=\left(\frac{\partial^2}{\partial t^2}-\frac{\partial^2}{\partial s^2}\right) 
\log \tau_n\label{2DTL2-sol}\\
&&\hspace{1.2cm} =\frac{\tau_{n+1}\tau_{n-1}}{\tau_n^2}-1\,,\nonumber\\
&&\tau_n=\det \left(\delta_{ij}+\frac{\phi_i \psi_j}{p_i-q_j}\right)_{_{1\leq i,j\leq N}}\nonumber\\
&&\quad     
=
\det \left(\delta_{ij}+\frac{\phi_j \psi_j}{p_i-q_j}\right)_{_{1\leq i,j\leq N}},\nonumber\\
     &&\phi_i=
      p_i^ne^{\frac{1}{2}(p_i-p_i^{-1})t+\frac{1}{2}(p_i+p_i^{-1})s+\phi_i^{(0)}}\,, \nonumber\\
&&       \psi_i =q_i^{-n}
      e^{-\frac{1}{2}(q_i-q_i^{-1})t-\frac{1}{2}(q_i+q_i^{-1})s+\psi_i^{(0)}}\,, \nonumber \\
&& \quad (i=1,2,\cdots,N)\,,\nonumber
\end{eqnarray}
where $p_i, q_i$ are positive constants and 
$\phi_i^{(0)}, \psi_i^{(0)}$ are real constants. 

Applying the delay reduction 
\begin{equation}
V_n(t,s)=w(z,s)\,, \qquad z=t+hn\,,
\end{equation}
to the 2DTL equation \eqref{2DTL2}, 
we obtain the delay 2DTL equation
\begin{equation}
\left(\frac{\partial^2}{\partial z^2}-\frac{\partial^2}{\partial s^2}\right) 
\log(1+w(z,s))=w(z+h,s)-2w(z,s)+w(z-h,s)\,,\label{delay2dtoda2}
\end{equation}
which can be written as
\begin{equation}
 \frac{\partial^2 r(z,s)}{\partial z^2}-\frac{\partial^2 r(z,s)}{\partial s^2}
=e^{r(z+h,s)}-2e^{r(z,s)}+e^{r(z-h,s)}
\end{equation} 
or
\begin{equation}
 \frac{\partial^2 u(z,s)}{\partial z^2}-\frac{\partial^2 u(z,s)}{\partial s^2}
=e^{u(z-h,s)-u(z,s)}-e^{u(z,s)-u(z+h,s)}
\end{equation} 
via the dependent variable transformation $r(z,s)=u(z,s)-u(z+h,s)=\log(1+w(z,s))$. 
Note that this delay 2DTL equation was considered by Villarroel and Ablowitz, 
and they found the Lax pair and established the inverse scattering transform\cite{ablowitz}.  
The delay 2DTL equation \eqref{delay2dtoda2} is transformed into the delay bilinear equation
\begin{equation}
(D_z^2-D_s^2)f(z,s)\cdot f(z,s)=2(f(z+h,s)f(z-h,s)-f(z,s)^2)
\end{equation}
via the dependent variable transformation
\begin{equation}
w(z,s)=\left(\frac{\partial^2}{\partial z^2}-\frac{\partial^2}{\partial s^2}\right) 
\log f(z,s)
=\frac{f(z+h,s)f(z-h,s)}{f(z,s)^2}-1\,.
\end{equation} 

We impose the reduction condition
\begin{equation}
 \log p_i- \log q_i=\frac{h}{2}(p_i-p_i^{-1}-q_i+q_i^{-1})\label{delay2dtodacond2}
\end{equation}
to the 
parameters $p_i, q_i$ $(i=1,2, \cdots , N)$ in the $N$-soliton solution \eqref{2DTL2-sol}. 
Then, by setting $z=x+hn$, we have
\begin{eqnarray}
&&\hspace{-0.8cm}\phi_j\psi_j=p_j^ne^{\frac{1}{2}(p_j-p_j^{-1})t+\frac{1}{2}(p_j+p_j^{-1})s
+\phi_j^{(0)}}q_j^{-n}
      e^{-\frac{1}{2}(q_j-q_j^{-1})t-\frac{1}{2}(q_j+q_j^{-1})s+\psi_j^{(0)}}\nonumber\\
&&=e^{\frac{1}{2}(p_j-p_j^{-1}-q_j+q_j^{-1})t+n(\log p_j-\log q_j)
+\frac{1}{2}(p_j+p_j^{-1}-q_j-q_j^{-1})s+\phi_j^{(0)}+\psi_j^{(0)}}\nonumber\\
&&=e^{\frac{1}{2}(p_j-p_j^{-1}-q_j+q_j^{-1})(t+hn)+
\frac{1}{2}(p_j+p_j^{-1}-q_j-q_j^{-1})s+\phi_j^{(0)}+\psi_j^{(0)}}\nonumber\\
&&=e^{\frac{1}{2}(p_j-p_j^{-1}-q_j+q_j^{-1})z+\frac{1}{2}(p_j+p_j^{-1}-q_j-q_j^{-1})s+\phi_j^{(0)}+\psi_j^{(0)}}\,.
\end{eqnarray}

Thus, by imposing the reduction condition \eqref{delay2dtodacond2} and 
setting $z=t+hn$, $f(z,s)=\tau_n(t,s)$, 
we obtain the following $N$-soliton solution of the delay 2DTL equation \eqref{delay2dtoda2}:
\begin{eqnarray}
&&w(z,s)=\left(\frac{\partial^2}{\partial z^2}-\frac{\partial^2}{\partial s^2}\right) 
\log f(z,s)\nonumber\\
&&\hspace{1.0cm} =\frac{f(z+h,s)f(z-h,s)}{f(z,s)^2}-1\,,\\
 && f(z,s)
=\det \left(\delta_{ij}+\frac{\Phi_j(z,s)}{p_i-q_j}\right)_{_{1\leq i,j\leq N}}\,,
\nonumber\\
     &&\Phi_j(z,s)=e^{\frac{1}{2}(p_j-p_j^{-1}-q_j+q_j^{-1})z
+\frac{1}{2}(p_j+p_j^{-1}-q_j-q_j^{-1})s+\Phi_j^{(0)}}\,,\nonumber
\end{eqnarray}
where $p_i, q_i$ must satisfy \eqref{delay2dtodacond2} 
and $\Phi_i^{(0)}$ are real constants.

By imposing another delay reduction 
\begin{equation}
V_n(t,s)=w(z_1,z_2)\,, \qquad z_1=t+h_1n\,,\qquad z_2=s+h_2n\,,
\end{equation}
to the 2DTL equation \eqref{delay2dtoda2}, where 
$h_1, h_2$ are nonzero real constants, 
we can also consider another delay 2DTL equation
\begin{eqnarray}
&&\left(\frac{\partial^2}{\partial z_1^2}-\frac{\partial^2}{\partial z_2^2}\right)
\log(1+w(z_1,z_2))=w(z_1+h_1,z_2+h_2)\nonumber\\ 
&& \qquad 
-2w(z_1,z_2)+w(z_1-h_1,z_2-h_2)\,,\label{delay2dtoda3}
\end{eqnarray}
which can be written as
\begin{equation}
 \frac{\partial^2 r(z_1,z_2)}{\partial z_1^2}-\frac{\partial^2 r(z_1,z_2)}{\partial z_2^2}
=e^{r(z_1+h_1,z_2+h_2)}-2e^{r(z_1,z_2)}+e^{r(z_1-h_1,z_2-h_2)}
\end{equation} 
or
\begin{eqnarray}
&& \frac{\partial^2 u(z_1,z_2)}{\partial z_1^2}-\frac{\partial^2 u(z_1,z_2)}{\partial z_2^2}
\nonumber\\
&&\quad =e^{u(z_1-h_1,z_2-h_2)-u(z_1,z_2)}-e^{u(z_1,z_2)-u(z_1+h_1,z_2+h_2)}
\end{eqnarray}
via the dependent variable transformation $r(z_1,z_2)=u(z_1,z_2)-u(z_1+h_1,z_2+h_2)
=\log(1+w(z_1,z_2))$. 
In this case, we can obtain the $N$-soliton solution by imposing the reduction condition
\begin{equation}
 \log p_i- \log q_i=\frac{h_1}{2}(p_i-p_i^{-1}-q_i+q_i^{-1})
+\frac{h_2}{2}(p_i+p_i^{-1}-q_i-q_i^{-1})\label{delay2dtodacond3}
\end{equation}
to the 
parameters $p_i, q_i$ $(i=1,2, \cdots , N)$ in the $N$-soliton solution \eqref{2DTL2-sol}. 
The $N$-soliton solution of the delay 2DTL equation \eqref{delay2dtoda3} is given as follows: 
\begin{eqnarray}
&&w(z_1,z_2)=\left(\frac{\partial^2}{\partial z_1^2}-\frac{\partial^2}{\partial z_2^2}\right) 
\log f(z_1,z_2)\nonumber\\
&&\hspace{1.0cm} =\frac{f(z_1+h_1,z_2+h_2)f(z_1-h_1,z_2-h_2)}{f(z_1,z_2)^2}-1\,,\\
 && f(z_1,z_2)
=\det \left(\delta_{ij}+\frac{\Phi_j(z_1,z_2)}{p_i-q_j}\right)_{_{1\leq i,j\leq N}}\,,
\nonumber\\
     &&\Phi_j(z_1,z_2)=e^{\frac{1}{2}(p_j-p_j^{-1}-q_j+q_j^{-1})z_1
+\frac{1}{2}(p_j+p_j^{-1}-q_j-q_j^{-1})z_2+\Phi_j^{(0)}}\,,\nonumber
\end{eqnarray}
where $p_i, q_i$ must satisfy \eqref{delay2dtodacond3} and $\Phi_i^{(0)}$ are real constants.
In the case of $h_2=0$, this delay 2DTL equation \eqref{delay2dtoda3} 
is reduced to the delay 2DTL equation \eqref{delay2dtoda2}, but 
it becomes another delay 2DTL equation in the case of $h_1=0$.  

Now we consider the reduction to the one-dimensional Toda lattice equation\cite{Levi}
\begin{equation}
\frac{d^2}{d z^2}
\log(1+w(z))=w(z+h)-2w(z)+w(z-h)\,,\label{delay1dtoda2}
\end{equation}
which can be written as
\begin{equation}
\frac{d^2r(z)}{dz^2}=e^{r(z+h)}-2e^{r(z)}+e^{r(z-h)} \,.
\end{equation}
or 
\begin{equation}
\frac{d^2u(z)}{dz^2}=e^{u(z-h)-u(z)}-e^{u(z)-u(z+h)} \,.
\end{equation}
by the transformnation $r(z)=u(z)-u(z+h)=\log(1+w(z))$. 
The delay 1DTL equation \eqref{delay1dtoda2} is transformed 
into the delay bilinear equation
\begin{equation}
D_z^2f(z)\cdot f(z)=2(f(z+h)f(z-h)-f(z)^2)
\end{equation}
via the dependent variable transformation
\begin{equation}
w(z)=\frac{d^2}{d z^2}
\log f(z)
=\frac{f(z+h)f(z-h)}{f(z)^2}-1\,.
\end{equation} 
We note that only 1-soliton solution survives under this reduction.  
Applying 
\begin{equation}
q_1=\frac{1}{p_1} 
\end{equation}
to the 1-soliton solution of the delay 2DTL equation \eqref{delay2dtoda2}, we obtain
\begin{eqnarray}
&&w(z)=\frac{d^2}{d z^2} \log f(z)=\frac{f(z+h)f(z-h)}{f(z)^2}-1\,,\\
 && f(z)
=1+\frac{e^{(p_1-p_1^{-1})z+\Phi^{(0)}}}{p_1-p_1^{-1}}\,,
\nonumber
\end{eqnarray}
where $p_1$ must satisfy 
\begin{equation}
2\log p_1=h(p_1-p_1^{-1})\,.
\end{equation}

\section{Conclusions}

We have considered delay reductions of the 2DTL equation and obtained 
the $N$-soliton solution of the delay analogues of the 2DTL equation. 
To the best of our knowledge, 
this is the first time that multi-soliton solutions of 
delay-differential equations has been obtained. 
We believe that our result is useful to study delay-differential equations.
In our forthcoming paper, we will propose a systematic method to generate 
delay soliton equations having multi-soliton solutions\cite{nakata}.

\section*{Declaration of competing interest}

The authors declare that they have no known competing financial interests or personal relationships that could have appeared to influence the work reported in this paper.

\section*{Acknowledgment}

The authors thank Prof. Willox for stimulating discussion. 
This work was partially supported by JSPS KAKENHI Grant Numbers 18K03435 and JST/CREST.

\end{document}